\documentclass[letterpaper]{article} 
\usepackage{aaai25} 
\usepackage{times}  
\usepackage{helvet}  
\usepackage{courier}  
\usepackage[hyphens]{url}  
\usepackage{graphicx} 
\usepackage{natbib}  
\usepackage{caption} 
\usepackage{algorithm}
\usepackage{algorithmic}

\usepackage{newfloat}
\usepackage{listings}
\DeclareCaptionStyle{ruled}{labelfont=normalfont,labelsep=colon,strut=off} 
\floatstyle{ruled}
\newfloat{listing}{tb}{lst}{}
\floatname{listing}{Listing}
\usepackage{xcolor}

\title{Outsourcing an Information Operation:\\A Complete Dataset of Tenet Media's Podcasts on Rumble}
\author {
    Laura Kurek\textsuperscript{\rm 1},
    Kevin Zheng\textsuperscript{\rm 1},
    Eric Gilbert\textsuperscript{\rm 1},
    Ceren Budak\textsuperscript{\rm 1}
}
\affiliations {
    \textsuperscript{\rm 1}University of Michigan\\
    lkurek@umich.edu, kyzheng@umich.edu, cbudak@umich.edu, eegg@umich.edu
}

\begin{document}

\maketitle

\begin{abstract}
Tenet Media, a U.S.-based, right-wing media company, hired six established podcasters to create content related to U.S. politics and culture during the 2024 U.S. presidential election cycle. After publishing content on YouTube and Rumble for nearly a year, Tenet Media was declared by the U.S. government to be funded entirely by Russia---making it effectively an outsourced state-sponsored information operation (SSIO). We present a complete dataset of the 560 podcast videos published by the Tenet Media channel on the video-sharing platform Rumble between November 2023 and September 2024. Our dataset includes video metadata and user comments, as well as high-quality video transcriptions, representing over 300 hours of video content. This dataset provides researchers with material to study a Russian SSIO, and notably on Rumble, which is an understudied platform in SSIO scholarship.
\end{abstract}

%
\begin{links}
    \link{Dataset}{https://zenodo.org/records/14629410}
\end{links}

\section{Introduction}
On September 4, 2024, the U.S. Department of Justice (DOJ) announced that the Russian Federation was again attempting to influence a U.S. general election~\cite{tucker_us_2024}. Part of this effort involved Tenet Media, a U.S.-based media company which the DOJ alleged had received over \$10 million from the Russian government to create podcast video content~\cite{doj_two_2024}. Beginning in late 2023, Tenet Media cultivated an internet presence across various platforms, including YouTube and Rumble, an alternative video hosting platform. Within a day of the DOJ indictment, Google removed Tenet Media channel from its YouTube platform~\cite{vynck_youtube_2024}. Rumble, meanwhile, has allowed the Tenet Media channel to remain on its site, though the channel has ceased uploading new content.

The Tenet Media channel published its first video on Rumble with the title: \textit{Tenet Media: Uncensored, Unapologetic, Unafraid}. Set to upbeat electronic music, the one-minute video introduced each of the six podcasters who were slated to create content for Tenet Media: Matt Christiansen, Tayler Hansen, Benny Johnson, Tim Pool, Dave Rubin, and Lauren Southern. The video description promised: ``Fearless voices live here. Coming November 6."

By funding these six U.S.-focused podcasters and providing them with editorial direction, the Russian government appears to have outsourced an information operation. Rather than creating fictional online personas, which it has done previously~\cite{golovchenko_cross-platform_2020}, Russia identified established content creators, who had already amassed sizable online followings among a target audience---in this case the U.S. electorate.

The Tenet Media channel ultimately published 560 videos on the Rumble platform between November 2023 and September 2024. These Tenet Media videos represent the content of a state-sponsored information operation (SSIO). In the study of SSIOs, ground truth datasets are scarce, resulting in an over-concentration of research on a small set of datasets, most notably the 2016 dataset of Russia's Internet Research Agency (IRA) ``troll" accounts, which were identified by Twitter and released to the public through Congressional action~\cite{fandos_house_2017}. 

\textbf{Contributions:} Totaling over 302 hours of video content, we present a dataset of metadata and transcripts for the 560 podcast videos published by the Tenet Media channel on Rumble. Our dataset serves to archive Tenet Media's online content, as its YouTube channel was promptly removed following the DOJ indictment and it is unclear how long this content will remain on Rumble. Given the Russian government's funding of Tenet Media, this work represents a rare, SSIO-related dataset. In addition to being new, it also differs from many SSIO-related datasets, in that is focused on paid content creators, as opposed to fake online accounts, and that it is a collection of videos, as opposed to social media posts. With it, researchers may study the impacts of foreign state funding on content creation. 
\begin{itemize}
    \item \textbf{Metadata:} Our dataset contains video details, including title, publication date, user engagement metrics. We also have collected user comments and related comment data.
    \item \textbf{Transcription:} Our dataset also contains high quality transcripts, generated by a state of the art speech-to-text model and manually checked for fidelity.
\end{itemize}

\section{Related Work}
\subsection{State-sponsored information operations} An information operation describes an organized attempt to manipulate the information environment towards a strategic goal~\cite{polychronis_working_2023, starbird_disinformation_2019}. State actors have leveraged the internet to conduct information operations anonymously and at scale~\cite{bradshaw_global_2019}. Previous ICWSM scholarship has analyzed the behavior of SSIOs online, including their use of images and deceptive language~\cite{zannettou_characterizing_2020,addawood_linguistic_2019}.

Russia has conducted multiple SSIOs in the internet era, targeting various international audiences, including the U.S.~\cite{jankowicz_how_2020}. Much of the Russian SSIO scholarship to date has focused on fake social media accounts and their impostor activity online~\cite{arif_acting_2018,lukito_coordinating_2020,linvill_troll_2020, golovchenko_cross-platform_2020}. This is due in part to the scarcity of SSIO-related datasets, concentrating the scholarship around a select few datasets, most notably the 2016 dataset of Russia's Internet Research Agency (IRA) ``troll'' accounts, which was released to the public~\cite{frommer_twitters_2017}. Our dataset provides an opportunity for SSIO researchers to study paid content creators, as opposed to fake ``troll'' accounts, as well as analyze podcast video content, instead of social media posts.

\subsection{Podcasts} Podcasts consist of episodic content distributed via the internet and often feature a recurring host and consistent focus, i.e., political punditry, life advice, etc.~\cite{sharon_peeling_2023}. The six Tenet Media-affiliated content creators produce episodic podcast videos with recurring hosts and a focus on U.S. politics and culture. Over the past two decades, a body of scholarship has sought to define what podcasts are, track their increasing popularity, and understand their impacts on the media environment, given their similarities to radio shows but with lower barriers to entry~\cite{berry_will_2006, pew_audio_2023, marchal_generative_2024}. In addition, there have been efforts to compile large-scale podcast datasets to aid researchers in this space~\cite{balci_dataset_2024,clifton_100000_2020,litterer_mapping_2024}.

\subsection{Rumble} Created in 2013, Rumble has emerged as a YouTube alternative---hosting content creators who have previously been ousted by other websites for violating platform policies. Viewership on Rumble grew precipitously during the COVID-19 pandemic, expanding from 1 million to 30 million users between 2020 and 2021~\cite{harwell_rumble_2021}. Previous research has found Rumble has attracted primarily right-wing content creators and similarly-minded audiences~\cite{stocking_role_2022, balci_podcast_2024}. Scholarship on Rumble overall is nascent, and to our knowledge there are no current dataset papers related to SSIOs and Rumble.

\section{Data Collection}
We developed an automated scraper which systematically moves through a Rumble channel to collect all videos chronologically, from the most recently published video to the first published video. The scraper navigates through the thumbnail grid on a channel's page and clicks into each video to collect additional metadata, including video comments. The scraper continues until no additional channel video pages are encountered. We collected a complete dataset of all 560 videos published by the Tenet Media channel between November 2023 and September 2024.

Data collection was conducted in December 2024. The Tenet Media channel published its last videos on September 4, 2024, following the DOJ indictment released on the day. The channel has been dormant since then. We note that the video engagement metrics (likes, views, etc.) reflect data collected in December 2024.

\begin{table*}[h]
\centering
\begin{tabular}{lll|lll}
\multicolumn{3}{l|}{\textbf{Video Title Named Entities}} & \multicolumn{3}{l}{\textbf{Video Description Named Entities}} \\ \hline
Entity                        & \#    & Label            & Entity                         & \#          & Label          \\ \hline
Matt Christiansen             & 90    & PERSON           & Matt Christiansen              & 207         & PERSON         \\
Dave Rubin \& Isabel Brown    & 89    & PERSON           & Dave Rubin                     & 132         & PERSON         \\
Lauren Southern               & 68    & PERSON           & Chris Gard                     & 114         & PERSON         \\
Tayler Hansen                 & 44    & PERSON           & Trump                          & 93          & PERSON         \\
Tim Pool                      & 38    & PERSON           & Biden                          & 87          & PERSON         \\
Biden                         & 21    & PERSON           & Brown                          & 85          & PERSON         \\
Benny Johnson                 & 19    & PERSON           & Lauren Southern                & 50          & PERSON         \\
The Culture War               & 16    & EVENT            & Tim Pool                       & 45          & PERSON         \\
Trump                         & 11    & PERSON           & Connect                        & 44          & PERSON         \\
Canada                        & 9     & GPE              & Tayler Hansen                  & 42          & PERSON         \\
Benny Johnson \&              & 8     & ORG              & Benny Johnson                  & 41          & PERSON         \\
Kamala                        & 7     & PERSON           & Matt Christiansen Hour         & 29          & PERSON         \\
Dave Rubin                    & 7     & PERSON           & Texas                          & 29          & GPE            \\
The Arena                     & 7     & WORK\_OF\_ART    & America                        & 29          & GPE            \\
The MC Hour                   & 6     & WORK\_OF\_ART    & FBI                            & 29          & ORG            \\
America                       & 6     & GPE              & Joe Biden                      & 23          & PERSON         \\
FBI                           & 6     & ORG              & California                     & 23          & GPE            \\
Supreme Court                 & 6     & ORG              & Lisa Elizabeth                 & 22          & PERSON         \\
Frank                         & 5     & PERSON           & Isabel Brown                   & 21          & PERSON         \\
Joe Biden                     & 5     & PERSON           & Chicago                        & 20          & GPE           
\end{tabular}
\caption{Top 20 named entities in video titles and descriptions.}
\label{Table 1}
\end{table*}

\section{Dataset Curation} 

\subsection{Metadata}
Our dataset contains comprehensive metadata for each Rumble video, including video comments and related comment metadata.

\vspace{8pt}
\textbf{Collected video metadata:}
\begin{itemize}
    \item \textbf{video\_id:} Unique identifier for the video, which is also located in the video URL.
    \item \textbf{video\_url:} URL for video on rumble.com.
    \item \textbf{video\_title:} Title of the video.
    \item \textbf{video\_host:} Video host, parsed from video title or tags.
    \item \textbf{video\_duration:} Duration of the video in string format.
    \item \textbf{video\_duration\_s:} Duration of the video in seconds. 
    \item \textbf{video\_thumbnail\_url:} URL to the video thumbnail, to allow researchers to study the images used to promote each video. Note that as an internal URL, this link may eventually expire.
    \item \textbf{video\_source\_url:} URL to the video source page, to allow researchers to easily download the video and view without having to log into ``rumble.com." Note that as an internal URL, this link may eventually expire.
    \item \textbf{video\_date:} Publication date of the video.
    \item \textbf{video\_views:} Total count of video views at the time of data collection.
    \item \textbf{upvotes\_count:} Total count of video upvotes, i.e., positive ratings, at the time of data collection.
    \item \textbf{downvotes\_count:} Total count of video downvotes, i.e., negative ratings, at the time of data collection.
    \item \textbf{video\_description:} Summary of the video as provided by the channel. 
    \item \textbf{video\_description\_cont:} Extended summary of the video, if present. Similar to~\cite{balci_dataset_2024}, we find that Rumble divides the video description in two segments, with the latter behind a `Show more' button.
    \item \textbf{video\_tags:} Keywords provided by the channel, which summarize the main topics and featured guests in the podcast.
    \item \textbf{video\_comment\_number:} Total count of video comments at the time of data collection.
\end{itemize}

\vspace{8pt}
\textbf{Collected comment metadata:}
For each video, we collected top-level comments and related data, listed below. We also captured whether the comment was low-scored. On Rumble, comments with a negative score (i.e., more downvotes than upvotes) are by default hidden. To view a hidden comment, users must click a ``Show low scored comment" button. With our scraper, we were able to parse whether or not a comment was hidden by default and thus low-scored. In the current iteration of this dataset, we did not collect comment replies to scope the collection task, but we can collect and release these in a future update if there is interest. 
\begin{itemize}
    \item \textbf{comment\_index:} Comment index.
    \item \textbf{username\_hash:} Username of the Rumble account which posted the comment. Usernames have been hashed for privacy considerations.
    \item \textbf{comment\_text:} The comment itself.
    \item \textbf{num\_likes:} Number of likes the comment had received at the time of data collection. This number can be negative, indicating dislikes.
    \item \textbf{num\_replies:} Number of replies the comment had received at the time of data collection. Note we did not collect replies to comments in this dataset.
    \item \textbf{low\_scored:} Boolean (true/false) indicating whether the comment was low scored or not.
\end{itemize}

\begin{figure*}[h]
\centering
\includegraphics[width=0.95\textwidth]{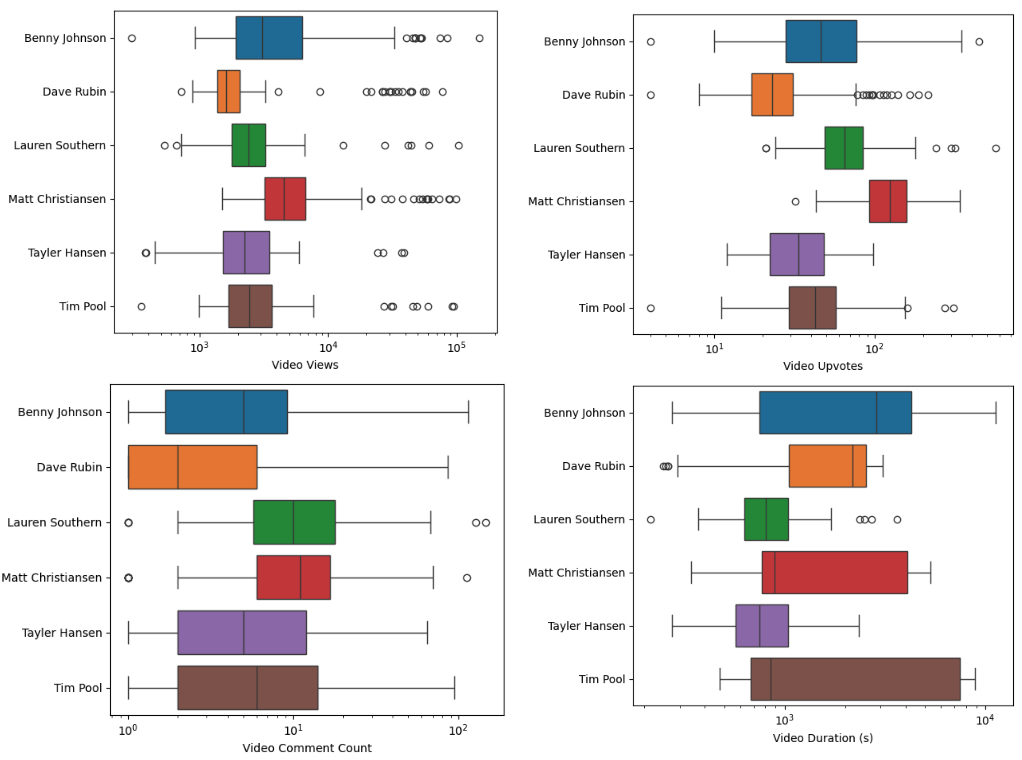}
\caption{Distribution of user engagement and video duration for each of the Tenet Media podcast hosts.}
\end{figure*}

\subsection{Transcription}
To extract what was said in the Rumble podcast videos--totaling over 300 hours of audio--we employed Whisper CPP,\footnote{\url{https://github.com/ggerganov/whisper.cpp}} a light-weight implementation of the automatic speech recognition model originally developed by OpenAI~\cite{radford_robust_2023}. We utilized the Whisper ``large-v2" model, a multi-lingual, 1500M-parameter model trained for 2.5x additional epochs more than the Whisper ``large" model.\footnote{\url{https://huggingface.co/openai/whisper-large-v2}} Given the tractable size of our dataset, we opted to use the largest of the Whisper models to produce the highest possible quality transcriptions. Generating these transcripts employed both CPU compute and NVIDIA V100 GPUs with 16GB on our university's internal computing cluster, totaling over 5 compute hours.

\subsubsection{Post-Processing of Transcripts}
The first author conducted a review of the each of the 560 transcriptions to ensure fidelity to the podcast videos. Of the 560 Whisper-generated transcripts, approximately 480 were assessed as high quality; for the remaining 80 transcripts, we took several post-processing steps to improve their quality. We first checked a sample of videos in their entirety: watching the entire video and following along in the transcript. We found that when a transcript correctly captured the podcast introduction, and had accurate punctuation and capitalization, it tended to be accurate throughout. From this fine-grain review, we observed that the Whisper ``large-v2" model often provided effective punctuation to denote pauses in conversation and correctly spelled less common words, such as podcast guests' social media handles.

Following a fine-grain check on a sample of transcripts, we then performed a course-grain check of all transcripts: confirming whether the transcript had punctuation, capitalization, and included introductory and closing remarks. In this quality control step, 80 transcripts were assessed as being of poor quality: some transcripts were only partially generated, some did not generate at all, and some did not contain any punctuation or capitalization. In addition, the Whisper model occasionally hallucinated repeating phrases that were not in fact said in the podcast videos, which has been noted in previous work~\cite{koenecke_careless_2024,mittal_towards_2024}.

To identify transcripts with repeating phrases, we compared transcription segments (typically 3-5 seconds in length), as output by the Whisper model's \textit{--output-json-full} format. We also identified any transcripts that failed to generate fully, i.e., transcripts that did not end with closing phrases. Each of the six content creators for Tenet Media tended to close their videos with a similar catch phrase, e.g. Benny Johnson's ``It's your boy Benny." If the video did not contain closing remarks, as some videos were snippets of long interviews, we watched the end of video to confirm that the transcript was complete.

To generate higher quality transcripts, we found that trimming the intro audio at the start of each video---often consisting of no sound or introductory music---and then rerunning the audio file through Whisper vastly improved transcript accuracy. Especially for transcripts which initially had no punctuation or capitalization, the simple change of removing the introductory audio resulted a transcript with precise punctuation and capitalization. As needed, we made minor manual adjustments, such as trimming multiple ``yous" at the start and end of transcripts, which Whisper appears to generate when it encounters dead audio. 

Following these post-processing steps, we re-assessed 555 transcripts to be of high quality. For these five videos, which consisted primarily of protest footage, the transcripts are of decent quality, but not as accurate as other videos which featured mic-ed hosts and guests.

\subsubsection{Content Warning}
Several transcripts include vulgar language, slurs, and other hate speech. In addition, several podcast videos include footage from U.S. university or college protests, which involve aggressive encounters between law enforcement and protesters.

\subsection{Data Release \& FAIR Guiding Principles}
In our dataset of podcast videos from the Tenet Media channel on Rumble, we have structured the data in a machine-readable json format. In accordance with FAIR (Findable Accessible, Interoperable, Re-usable) guidelines, we have placed our dataset on the dataset-sharing service the Zenodo. With a digital object identifier (DOI), other researchers may discover and access this data. With a creative commons non-commercial license, other researchers are able to use and adapt this data freely. 

\begin{figure*}[h]
\centering
\includegraphics[width=0.90\textwidth]{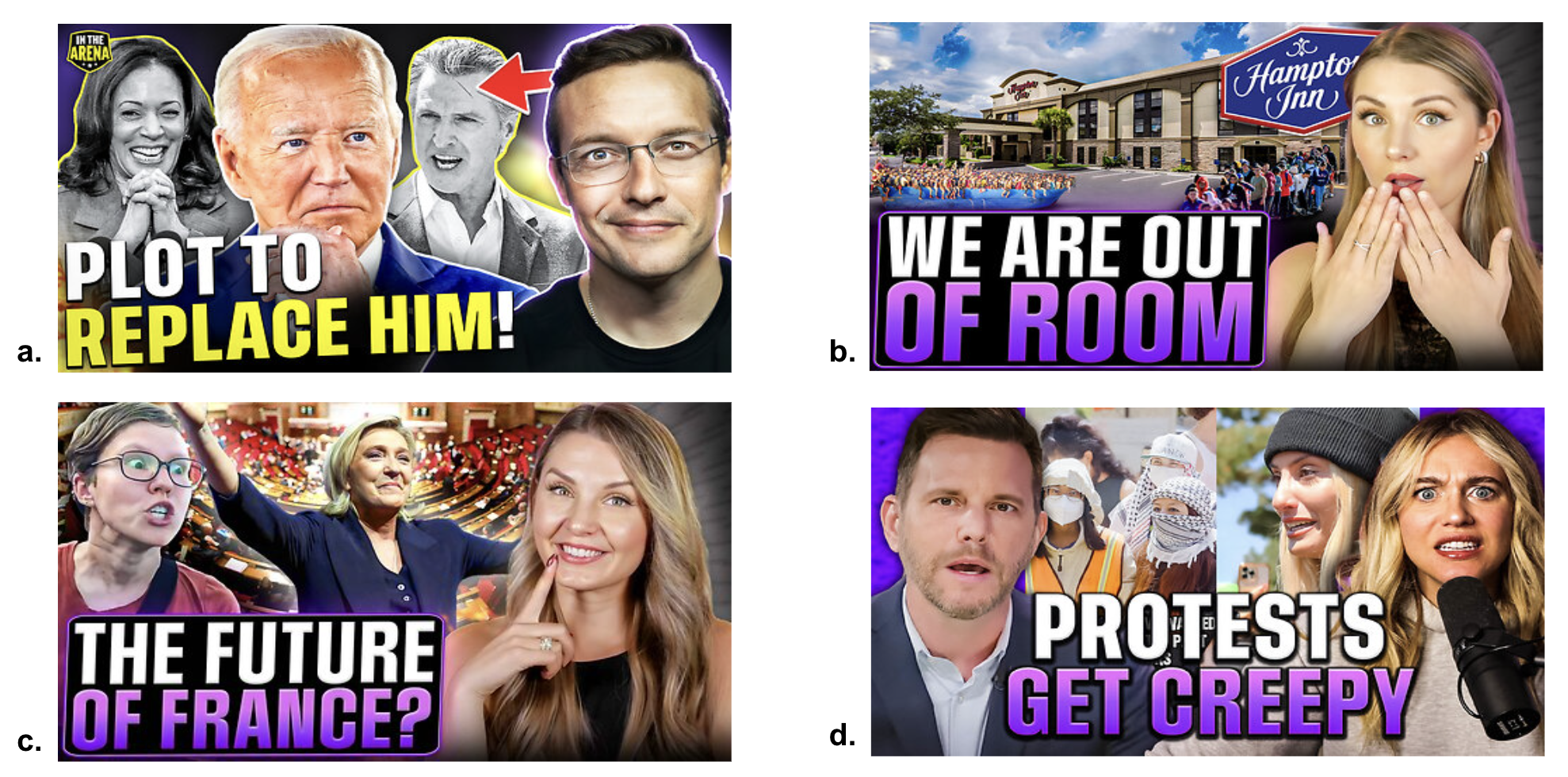}
\captionsetup{width=0.7\textwidth}
\caption{Tenet media videos with top user engagement: a) most viewed video, b) most commented on video, c) most upvoted video, and d) most downvoted video.}
\end{figure*}

\section{Preliminary Analysis}
We conducted a preliminary descriptive analysis of the video metadata, as well as generated transcripts. 

\textbf{Video duration}: The video duration across the 560 Tenet Media podcast videos totaled 302.8 hours. The average video has a duration of 32 minutes. The shortest video was 1 minute in length, and the longest was 3.13 hours long. Lauren Southern and Tayler Hansen tend to host shorter videos than the other four hosts. For example, Tayler Hansen frequently produced 15 minute `man-on-the-street' interviews, while Tim Pool often hosted 2-hour panel discussions with invited guests. 

\begin{table*}[t]
\centering
\begin{tabular}{rrl}
\multicolumn{1}{l}{\textbf{Topic}} & \textbf{Count}              & \textbf{Topic Words}                           \\ \hline
\multicolumn{1}{r|}{0}             & \multicolumn{1}{r|}{39}  & case, trump, court, question, law,             \\
\multicolumn{1}{r|}{}              & \multicolumn{1}{l|}{}    & tim, biden, guy, government, says              \\ \cline{3-3} 
\multicolumn{1}{r|}{1}             & \multicolumn{1}{r|}{20}  & gay, guys, white, media, different,            \\
\multicolumn{1}{r|}{}              & \multicolumn{1}{l|}{}    & gender, kids, culture, opinion, month          \\ \cline{3-3} 
\multicolumn{1}{r|}{2}             & \multicolumn{1}{r|}{103} & dave, love, years, guy, kids,                  \\
\multicolumn{1}{r|}{}              & \multicolumn{1}{l|}{}    & today, isabel, whats, laughing, guys           \\ \cline{3-3} 
\multicolumn{1}{r|}{3}             & \multicolumn{1}{r|}{11}  & black, israel, crowd, chicago, vote,           \\
\multicolumn{1}{r|}{}              & \multicolumn{1}{l|}{}    & money, white, riot, booing, reparations        \\ \cline{3-3} 
\multicolumn{1}{r|}{4}             & \multicolumn{1}{r|}{22}  & canada, government, canadian, canadians,       \\
\multicolumn{1}{r|}{}              & \multicolumn{1}{l|}{}    & housing, years, trudeau, covid, money, country \\ \cline{3-3} 
\multicolumn{1}{r|}{5}             & \multicolumn{1}{r|}{70}  & women, men, kids, school, white,               \\
\multicolumn{1}{r|}{}              & \multicolumn{1}{l|}{}    & years, guys, woman, world, work                \\ \cline{3-3} 
\multicolumn{1}{r|}{6}             & \multicolumn{1}{r|}{61}  & case, court, law, gun, crime,                  \\
\multicolumn{1}{r|}{}              & \multicolumn{1}{l|}{}    & police, amendment, federal, guns, judge        \\ \cline{3-3} 
\multicolumn{1}{r|}{7}             & \multicolumn{1}{r|}{53}  & trump, biden, joe, president, donald,          \\
\multicolumn{1}{r|}{}              & \multicolumn{1}{l|}{}    & country, kamala, party, man, harris            \\ \cline{3-3} 
\multicolumn{1}{r|}{8}             & \multicolumn{1}{r|}{36}  & congress, trump, house, president, committee,  \\
\multicolumn{1}{r|}{}              & \multicolumn{1}{l|}{}    & members, staff, office, donald, military      
\end{tabular}
\caption{Preliminary topic modeling analysis on the Tenet Media transcripts}
\label{Table 2}
\end{table*}

\textbf{User Engagement}: Our dataset include user engagement metrics, including views, number of comments, upvotes, and downvotes. In Figure 1, we include video thumbnails from the most engaged with videos on the Tenet Media channel.

\begin{itemize}
    \item \textbf{Video views:} The average video was viewed over 8,340 times. The least-viewed video received only 295 views. The most-viewed video, ``\textit{Will Biden Actually Be REPLACED?}", was from podcaster Benny Johnson and received over 151,000 views. Most videos uploaded receive in the range of 1,000 and 10,000 views. 
    \item \textbf{Comments:} The average video received around 12 comments, with some videos receiving no comments. The most commented-on video, ``\textit{NYC Hotels TAKEN OVER By Migrants}", was from podcaster Lauren Southern and received 147 comments. Most videos received fewer than 20 comments, with a few outlying videos receiving greater than 100 comments. 
    \item \textbf{Upvotes:} The average video received around 75 upvotes, with a minimum of 4 upvotes and a maximum of 568 upvotes. The most upvoted video was titled ``\textit{The Left DEFEATED in France}", and featured podcaster Lauren Southern. Most videos had fewer than 100 upvotes. Matt Christiansen's videos tended to receive more upvotes than videos from other hosts, with many videos receiving more than 100 upvotes. 
    \item \textbf{Downvotes:} The average video received fewer than 3 downvotes, with some videos receiving no downvotes. The most downvoted video,``\textit{Latest Palestine Protest Footage Is Very Creepy}", received 25 downvotes and featured podcaster Dave Rubin and his frequent guest Isabel Brown. Matt Christiansen and Tayler Hansen received very few downvotes, and Tim Pool received the most downvotes overall. 
\end{itemize}

\textbf{Content Descriptors:} For video titles, descriptions, and tags, we extracted the top 20 most frequent named entities. These content descriptors can provide insight into central topics across the podcast videos. Our results are summarized in Table 1.
\begin{itemize}
    \item \textbf{Video Titles:} Many of the top title entities are related to the names of six U.S. podcasters hired by Tenet Media and the names of their podcasts (e.g., ``The Culture War with Tim Pool").\footnote{The six podcasters/internet personalities hired by Tenet Media: Matt Christiansen, Tayler Hansen, Benny Johnson, Tim Pool, Dave Rubin, and Lauren Southern.} Beyond podcaster names, other frequent entities include ``Biden" (21), ``Trump" (11) and ``Kamala" (7), suggesting a focus around U.S. politics and the 2024 U.S. presidential election.
    \item \textbf{Video Descriptions:} The top video description entities are similarly related to the podcast hosts and their show names (e.g., ``Matt Christiansen Hour"). Other frequent description entities are ``Trump" (93) and ``Biden" (87), as well as ``Texas" (29), ``America" (29), ``California" (23) and ``Chicago" (20), reflecting a focus on the U.S. politics at both the federal and state levels.
    \item \textbf{Video Tags:} The top video tags entries are primarily related to the podcast hosts and their show names (e.g. ``Dave Rubin's People of the Internet")
\end{itemize}

\textbf{Transcript Analysis:} With the transcripts generated from the podcast videos, we conducted preliminary topic modeling with BERTopic,\footnote{\url{https://maartengr.github.io/BERTopic/index.html}} which employs a sentence-transformers model and TF-IDP clustering to identify underlying topics in a corpus of documents. Our results are summarized in Table 2.

In addition to extracting named entities from video titles and descriptions, conducting topic modeling on the video transcripts provides further insight into common themes found in Tenet Media's content. In Topics 7 and 8, we again see focus on the 2024 U.S. presidential election and U.S. federal politics more broadly. 

Beyond national politics, we see a focus on gender and sexual orientation in Topics 1 and 5. From our initial review, several of the Tenet Media videos contained anti-LGBTQ and anti-trans rhetoric, in addition to promoting traditional gender norms. Topic 3 is likely related to protests and social movements in the U.S. In our initial review, Tayler Hansen made several videos attending protests in the U.S., including several pro-Palestine protests which took place across university campuses in 2024. Topic 4 is likely related to content from Lauren Southern, a right-wing commentator from Canada frequently critical of the Canadian Liberal Party.

\section{Conclusion}
We present a complete dataset of the 560 Rumble videos published by Tenet Media, a U.S.-based right-wing media company funded by the Russian government. Our dataset consists of video metadata, including user engagement metrics and user comments, as well as high-quality video transcripts. From a preliminary analysis on the dataset, content from the Tenet Media channel focused on U.S. national politics, including the 2024 presidential election, as well as promoted right-wing positions on social and cultural issues. Our dataset is a resource for researchers to investigate the impacts of foreign state funding on content creation.

\section{Limitations}
In the video metadata provided in our dataset, the user engagement metrics reflect data collected in December 2024. These metrics maybe not be fully accurate depending on when researchers work with the data in future. In addition, while we did our best to catch repetitive hallucinated phrases in the video transcripts, there still may be instances where transcriptions contain text that was not actually said in the Rumble video. This is an artifact of using the Whisper model for transcription, and it has been documented in prior work~\cite{koenecke_careless_2024,mittal_towards_2024}.

\section{Acknowledgments}
We extend thanks to the University of Michigan Great Lakes Cluster support team, who were incredibly helpful in assisting with our data pipeline configuration. We also extend thanks to Lavinia Dunagan for expert Matplotlib guidance.

\bibliography{references}


\section{Ethics Statement and Author Checklist}

\paragraph{Ethics Statement} The creation and release of a podcast dataset poses several ethical considerations. First, there is a the risk of mistranscribing someone’s speech in the podcast videos. To mitigate this risk, we conducted a comprehensive review of transcripts to minimize the prevalence of hallucinations. We re-generated and edited transcripts as needed. There still could be mistakes, however, and we encourage other researchers to take this into account when working with the data. Second, regarding privacy concerns, we have hashed commenter usernames before releasing the dataset, as regular users of Rumble have some expectation of privacy. We did not anonymize the Tenet Media podcast hosts or their guest names, which might be mentioned in video titles, descriptions, tags or transcripts. In their role as podcast hosts and guests, they are acting as public figures with fewer expectations of privacy. Third, there is a potential that this data could be used to train large language models. We again note here that the Tenet Media podcast videos contain vulgar language, slurs, and other hate speech. We urge researchers to consider whether to include such as dataset in training data for other applications.

\newcommand{\answerYes}[1]{\textcolor{blue}{#1}} 
\newcommand{\answerNo}[1]{\textcolor{teal}{#1}} 
\newcommand{\answerNA}[1]{\textcolor{gray}{#1}} 
\newcommand{\answerTODO}[1]{\textcolor{red}{#1}} 

\lstset{%
	basicstyle={\footnotesize\ttfamily},
	numbers=left,numberstyle=\footnotesize,xleftmargin=2em,
	aboveskip=0pt,belowskip=0pt,%
	showstringspaces=false,tabsize=2,breaklines=true}
\floatstyle{ruled}
\newfloat{listing}{tb}{lst}{}
\floatname{listing}{Listing}

\begin{enumerate}

\item For most authors...
\begin{enumerate}
    \item  Would answering this research question advance science without violating social contracts, such as violating privacy norms, perpetuating unfair profiling, exacerbating the socio-economic divide, or implying disrespect to societies or cultures?
    \answerYes{Yes, our dataset is focused on understanding a state-sponsored information operation (SSIO) on a public video-sharing platform.}
  \item Do your main claims in the abstract and introduction accurately reflect the paper's contributions and scope?
    \answerYes{Yes, we provide an overview of our dataset and its usefulness to researchers.}
   \item Do you clarify how the proposed methodological approach is appropriate for the claims made? 
    \answerYes{Yes, we discuss our data collection and cleaning approach in the Data Collection and Dataset Curation sections.}
   \item Do you clarify what are possible artifacts in the data used, given population-specific distributions?
    \answerNA{NA}
  \item Did you describe the limitations of your work?
    \answerYes{Yes, please see the Limitations section.}
  \item Did you discuss any potential negative societal impacts of your work?
    \answerYes{Yes, please see the Ethics Statement.}
      \item Did you discuss any potential misuse of your work?
   \answerYes{Yes, please see the Ethics Statement.}
    \item Did you describe steps taken to prevent or mitigate potential negative outcomes of the research, such as data and model documentation, data anonymization, responsible release, access control, and the reproducibility of findings?
    \answerYes{Yes, please see the Ethics Statement.}
  \item Have you read the ethics review guidelines and ensured that your paper conforms to them?
    \answerYes{Yes.}
\end{enumerate}

\item Additionally, if your study involves hypotheses testing...
\begin{enumerate}
  \item Did you clearly state the assumptions underlying all theoretical results?
    \answerNA{NA}
  \item Have you provided justifications for all theoretical results?
   \answerNA{NA}
  \item Did you discuss competing hypotheses or theories that might challenge or complement your theoretical results?
   \answerNA{NA}
  \item Have you considered alternative mechanisms or explanations that might account for the same outcomes observed in your study?
   \answerNA{NA}
  \item Did you address potential biases or limitations in your theoretical framework?
   \answerNA{NA}
  \item Have you related your theoretical results to the existing literature in social science?
   \answerNA{NA}
  \item Did you discuss the implications of your theoretical results for policy, practice, or further research in the social science domain?
    \answerNA{NA}
\end{enumerate}

\item Additionally, if you are including theoretical proofs...
\begin{enumerate}
  \item Did you state the full set of assumptions of all theoretical results?
    \answerNA{NA}
	\item Did you include complete proofs of all theoretical results?
    \answerNA{NA}
\end{enumerate}

\item Additionally, if you ran machine learning experiments...
\begin{enumerate}
  \item Did you include the code, data, and instructions needed to reproduce the main experimental results (either in the supplemental material or as a URL)?
    \answerNA{NA}
    \item Did you specify all the training details (e.g., data splits, hyperparameters, how they were chosen)?
    \answerNA{NA}
    \item Did you report error bars (e.g., with respect to the random seed after running experiments multiple times)?
    \answerNA{NA}
    \item Did you include the total amount of compute and the type of resources used (e.g., type of GPUs, internal cluster, or cloud provider)?
    \answerNA{NA}
     \item Do you justify how the proposed evaluation is sufficient and appropriate to the claims made? 
    \answerNA{NA}
     \item Do you discuss what is ``the cost`` of misclassification and fault (in)tolerance?
    \answerNA{NA}
  
\end{enumerate}

\item Additionally, if you are using existing assets (e.g., code, data, models) or curating/releasing new assets, \textbf{without compromising anonymity}...
\begin{enumerate}
  \item If your work uses existing assets, did you cite the creators?
    \answerNA{NA}
  \item Did you mention the license of the assets?
    \answerYes{Yes, please see Data Release \& FAIR Guiding Principles.}
  \item Did you include any new assets in the supplemental material or as a URL?
    \answerYes{Yes, we provide a Zenodo URL to our dataset on page 1.}
  \item Did you discuss whether and how consent was obtained from people whose data you're using/curating?
    \answerYes{Yes, please see the Ethics Statement.}
  \item Did you discuss whether the data you are using/curating contains personally identifiable information or offensive content?
    \answerYes{Yes, please see the Ethics Statement and Content Warning in the subsection Transcription under Data Curation.}
\item If you are curating or releasing new datasets, did you discuss how you intend to make your datasets FAIR (see \citet{wilkinson_fair_2016})?
\answerYes{Yes, please see Data Release \& FAIR Guiding Principles.}
\item If you are curating or releasing new datasets, did you create a Datasheet for the Dataset (see \citet{gebru_datasheets_2021})? 
\answerNo{No, as we consider this dataset paper to serve as a datasheet. We describe the motivation for collecting this data, what is in the data, and its limitations.}
\end{enumerate}

\item Additionally, if you used crowdsourcing or conducted research with human subjects, \textbf{without compromising anonymity}...
\begin{enumerate}
  \item Did you include the full text of instructions given to participants and screenshots?
    \answerNA{NA}
  \item Did you describe any potential participant risks, with mentions of Institutional Review Board (IRB) approvals?
    \answerNA{NA}
  \item Did you include the estimated hourly wage paid to participants and the total amount spent on participant compensation?
    \answerNA{NA}
   \item Did you discuss how data is stored, shared, and deidentified?
   \answerNA{NA}
\end{enumerate}

\end{enumerate}

\section{Appendices}

\subsection{Video Tags}
We extracted the top 20 most frequent named entities for video 
tags. They are primarily related to the podcast hosts and their show names.
\begin{table}[h]
\centering
\begin{tabular}{ll}
\multicolumn{2}{l}{\textbf{Tags}}   \\ \hline
Tag                           & \#  \\ \hline
tenet media                   & 350 \\
entertainment                 & 306 \\
news                          & 159 \\
dave rubin                    & 123 \\
matt christiansen             & 120 \\
tenet                         & 115 \\
isabel brown                  & 114 \\
reactions                     & 76  \\
lauren southern               & 74  \\
tim pool                      & 50  \\
the culture war               & 44  \\
matt christiansen hour        & 41  \\
tayler hansen                 & 40  \\
joe biden                     & 33  \\
benny johnson                 & 30  \\
people of the internet        & 30  \\
donald trump                  & 28  \\
podcasts                      & 27  \\
in the arena                  & 26  \\
the culture war with tim pool & 22 
\end{tabular}
\caption{Top 20 tag strings in video tags.}
\label{Table 3}
\end{table}

\end{document}